\documentclass{ws-procs9x6}
\usepackage{epsfig}
\begin{document}


\title{Photon production in relativistic nuclear collisions at SPS and RHIC
energies}

\author{Simon Turbide$^{1}$\footnote{\uppercase{W}ork presented at the
26th annual
\uppercase{M}ontreal-\uppercase{R}ochester-\uppercase{S}yracuse-\uppercase{T}oronto 
conference (\uppercase{MRST} 2004) on high energy physics, \uppercase{M}ontreal, 
\uppercase{QC}, \uppercase{C}anada, 12-14 \uppercase{M}ay 2004.} , 
Ralf Rapp$^{2}$, and  Charles Gale$^1$}

\address{$^1$ Department of Physics, McGill University, 3600 University
Street\\ Montreal, QC, H3A 2T8 Canada\\
 $^2$ Cyclotron Institute and Physics Department, Texas A\&M University\\ College
 Station, TX, 77843-3366 USA}

\maketitle

\abstracts{Chiral Lagrangians are used to compute the production
rate of photons from the hadronic phase of relativistic nuclear collisions. 
Special attention is paid to the role of the pseudovector $a_1$ meson. 
Calculations that include reactions with strange mesons, 
hadronic form factors and vector spectral densities consistent with 
dilepton production, as well as the emission from a 
quark-gluon plasma and primordial nucleon-nucleon collisions,
reproduce the photon spectra measured at the Super 
Proton Synchrotron (SPS). Predictions for the Relativistic Heavy Ion
Collider (RHIC) are made.
}

\section{Introduction}

The electromagnetic radiation emitted during relativistic heavy ion
collisions has been an intense subject of research for several years. 
As real and virtual photons essentially do not suffer final state 
interactions, they are good probes of the local condition in the hot 
matter created.  
One of the goals of this research program is to investigate whether a 
new phase of matter (generically called a Quark-Gluon Plasma (QGP)) is 
formed in nuclear collisions.  The existence of this state is a 
prediction of Quantum Chromodynamics (QCD). However, the measured 
photons cannot be tagged individually according to their emission
phase, and the QGP contribution therefore cannot be isolated in a 
direct fashion. In order to compare with data, one needs to compute the 
production rate of photons from the QGP phase, but also from the reactions
occurring after hadronization, and from a potential mixed phase. A previous 
study~\cite{KLS91} has shown a comparable brightness of partonic and 
hadronic phases.  Later works~\cite{xsb,Song93} have
highlighted the important role played by the $a_1$ pseudovector meson 
in the hadronized, non-perturbative sector.
Here, we reconsider the axial vector together with constraints imposed by 
known hadronic phenomenology. We discuss the role of form factors, 
of the QGP and of the Cronin efffect in primordial
nucleon-nucleon collisions in the interpretation of the WA98 photon
data, measured at the CERN SPS. Predictions for RHIC will also be shown.

\section{Hadronic phase}
\subsection{Massive Yang-Mills approach}

The interactions of hadrons can be described by Lagrangians which should
respect the fundamental symmetries of QCD. 
The introduction of vector mesons within the Massive Yang-Mills (MYM) 
approach provides an adequate meson phenomenology. A chiral U$(3)_L
\times$ U$(3)_R$ MYM Lagrangian can be written as\cite{Song93,gomm}:
\begin{eqnarray}
\mbox{\em L} &=& \textstyle{\frac{1}{8}} F_\pi^2 {\rm Tr} D_\mu U D^\mu
U^\dag + \frac{1}{8}
F_\pi^2 {\rm Tr} M (U + U^\dag -2)\nonumber \\& &  -
\textstyle{\frac{1}{2}}
{\rm Tr} \left(F_{\mu \nu}^L
{F^L}^{\mu \nu} + F_{\mu \nu}^R {F^R}^{\mu \nu} \right) + m_0^2 {\rm
Tr}
\left(A_\mu^L {A^L}^{\mu \nu} + A_\mu^R {A^R}^\mu\right)+ \nonumber \\ & & 
\gamma {\rm Tr}F_{\mu \nu}^L U F^{R \mu \nu}U^\dag
 -i\xi {\rm Tr}\left(D_\mu UD_\nu U^\dag F^{L \mu \nu}
+D_\mu U^\dag D_\nu U F^{R \mu \nu}\right)\, .
\label{Lmym}
\end{eqnarray}
In the above,
\begin{eqnarray}
\lefteqn{U = \exp \left( \frac{2 i}{F_\pi} \sum_i \frac{\phi_i
\lambda_i}{\sqrt{2}}\right) = \exp\left( \frac{2 i}{F_\pi}
\mbox{\boldmath $\phi$} \right)\ ,}\ \nonumber \\
 & & A_\mu^L = \textstyle{\frac{1}{2}}(V_\mu + A_\mu)\ , \nonumber \\
 & & A_\mu^R = \textstyle{\frac{1}{2}}(V_\mu - A_\mu)\ , \nonumber \\
 & & F_{\mu \nu}^{L, R}  = \partial_\mu A_\nu^{L, R} - \partial_\nu
A_{\mu}^{L, R} -
i g_0 \left[A_{\mu}^{L, R}, A_\nu^{L, R} \right]\ ,\nonumber \\
 & & D_\mu U = \partial_\mu U - i g_0 A_\mu^L U + i g_0 U A_\mu^R\
,\nonumber \\
 & & M = \frac{2}{3} \left[ m_K^2 + \frac{1}{2} m_\pi^2\right] -
\frac{2}{\sqrt{3}}
(m_K^2 - m_\pi^2) \lambda_8\ .
\end{eqnarray}

The vector and axial vector mesons are included in the theory as
gauge bosons.  The electromagnetic field is implemented through a U(1)
transformation, which leads to the well-known\cite{sak} vector-meson-dominance
(VMD):
\begin{equation}
L_{em}=-C m_\rho^{2} \rho^0_\mu B^\mu\ . 
\end{equation}

This theory leaves 5 unknown parameters $(C,m_0,g_0,\gamma,\xi)$ that
we can
fit using measured values of $\Gamma_{\rho\rightarrow e^-e^+}$, $m_{\rho}$,
$m_{a_1}$, $\Gamma_{\rho\rightarrow \pi\pi}$, $\Gamma_{a_1\rightarrow
\rho\pi}$.  This procedure still allows for two possible sets of solution,
\begin{eqnarray}
({\rm I}):\ \ \ \ \ \tilde{g} = 10.3063, \ \gamma = 0.3405,\
\xi = 0.4473,\ m_0 = 0.6253\ \mbox{ GeV}\ ;\nonumber \\
({\rm II}):\ \ \ \ \,
\tilde{g} = 6.4483,\ \gamma = - 0.2913,\ \xi=0.0585,\ m_0 = 0.875\
\mbox{ GeV}\ ,
\end{eqnarray}
where $\tilde{g} = g_0 / \sqrt{1 - \gamma}$.  To discriminate between
the two sets, one can take advantage of another hadronic measurement, 
the $D$- to $S$-wave ratio in the final state of $a_1\rightarrow \rho\pi$.  
In addition, the radiative decay width of the $a_1$ is known experimentally, 
albeit with somewhat less precision. Note also that as mesons are not 
fundamental fields, form factors need to be included in the theoretical 
estimates. This turns out to be especially important in the evaluation of 
strong off-shell vertices like the one contained in 
$\Gamma_{a_1 \rightarrow \pi\gamma}$\cite{RG99}. 
An evaluation of the $D/S$ ratio and of the radiative decay width yields
\begin{eqnarray}
&({\rm I}): &  D/S = 0.36,\ \Gamma_{a_1 \rightarrow \pi +\gamma} = 2.2
\mbox{ MeV}\
\nonumber \\
&({\rm II}):&  D/S = - 0.099,\ \Gamma_{a_1 \rightarrow \pi +\gamma} =
0.033 \mbox{ MeV}\   \  . 
\end{eqnarray}
The results with set II agree well with the experimental $D/S$ value 
of $- 0.107 \pm$ 0.016.  However, neither set I nor set II reproduces the
experimental radiative decay width, $\Gamma_{a_1 \rightarrow \pi\gamma}
$=0.64 $\pm$ 0.246 MeV.  For the sake of quantitative analysis,
we define a third set as 
\begin{eqnarray}
({\rm III}):\ \ \ \ \ \tilde{g} = 5.834, \ \gamma = -0.464,\
\xi = 0.1157,\ m_0 = 0.847 \mbox{ GeV}\ ,
\end{eqnarray}
leading to 
\begin{eqnarray}
m_{a_1}=1.4 \mbox{ GeV} ,\ m_{\rho}=0.7 \mbox{ GeV} ,\ \Gamma_{\rho}=0.17
\mbox{ GeV}\ \\
 \Gamma_{a_1}=0.3 \mbox{ GeV},\ D/S=-0.49,\ \Gamma_{a_1 \rightarrow \pi
+\gamma}=0.44 \mbox{ MeV}\ . 
\end{eqnarray}

\begin{figure}[ht!]
\psfig{figure=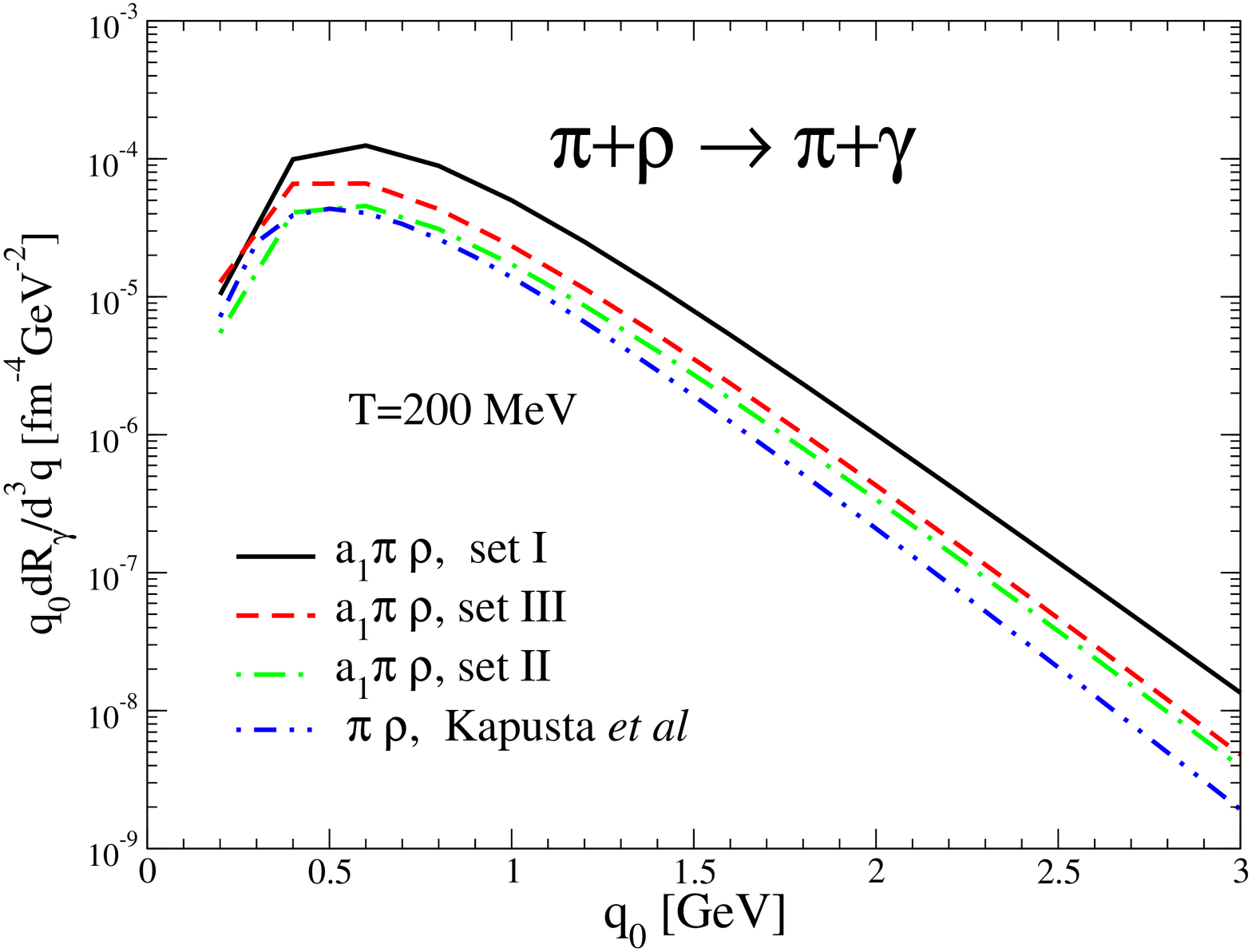,width=1.7in,angle=-90}
\psfig{figure=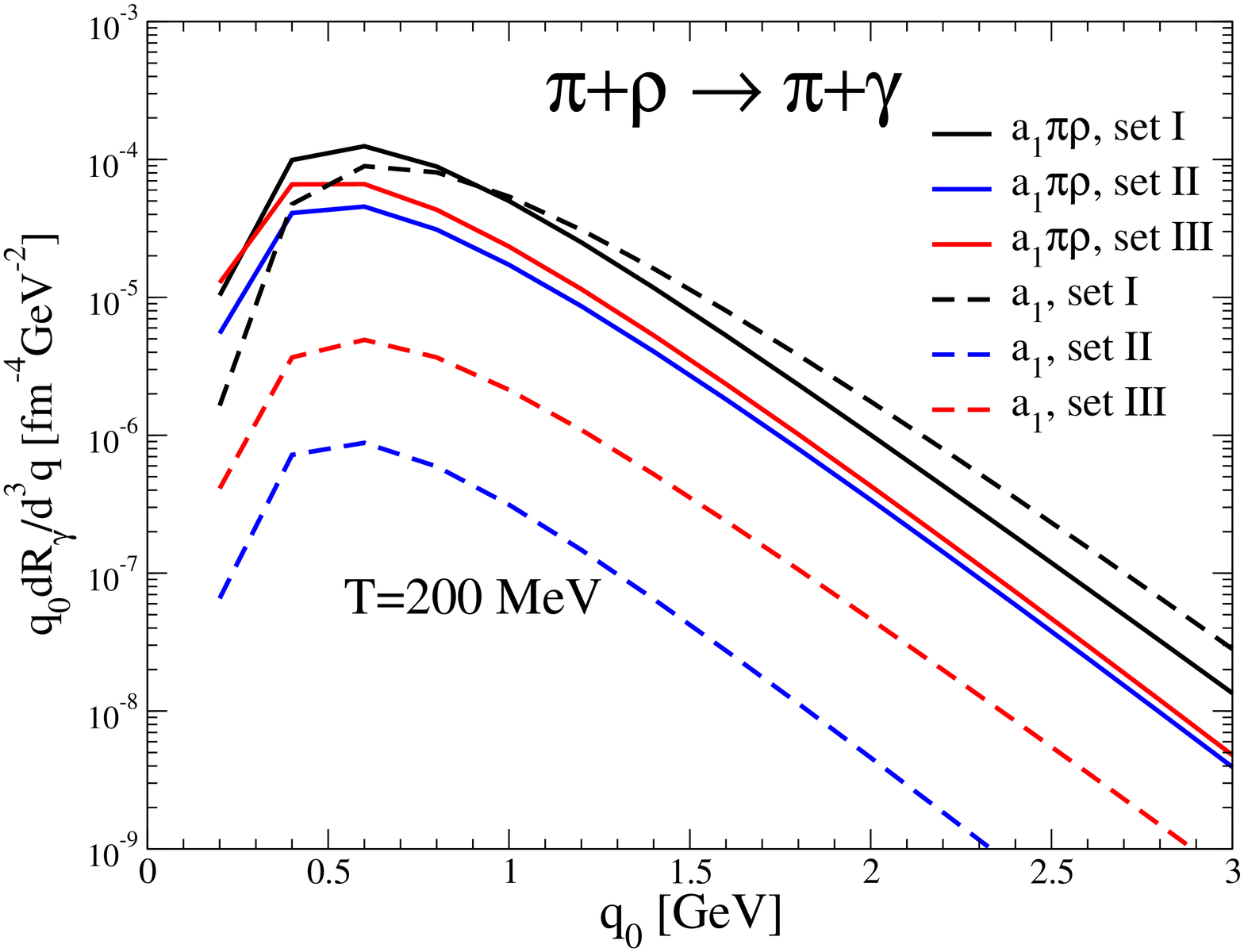,width=1.7in,angle=-90}
\caption{Left panel: $\pi+\rho \rightarrow \pi +\gamma$ from
$\pi\rho a_1$ intermediate states added coherently, at a temperature 
of T=200 MeV for the
three sets discussed in section I. The dash-double-dotted curve represents 
the result from Ref.~[\protect\refcite{KLS91}].  Right panel: $a_1$ and
$a_1\pi\rho$ contribution to $\pi+\rho \rightarrow \pi +\gamma$. No form factors
are included here, see next subsection.}
\end{figure}

\subsection{Photon production rates}
\subsubsection{The role of the $a_1$ pseudovector}

The photon production rate from hadronic matter is evaluated in relativistic
kinetic theory~\cite{KLS91,Song93},
\begin{eqnarray}
q_0 \frac{dR_\gamma}{d^3q} &=&\int \frac{d^3p_1}{2(2\pi)^3E_1}
\frac{d^3p_2}{2(2\pi)^3E_2}\frac{d^3p_3}{2(2\pi)^3E_3}(2\pi)^4
\delta^{(4)}(p_1+p_2-p_3-q)\nonumber \\ &&
\times \left|M_{12\to 3\gamma} \right|^2
\frac{f(E_1)f(E_2)[1\pm f(E_3)]}{2(2\pi)^3} \  ,
\label{rate_kin}
\end{eqnarray}
where $M_{12\to 3\gamma}$ is the process amplitude and $f$ is a  
Bose-Einstein distribution function. 
Figure 1 shows the reaction
$\pi+\rho \rightarrow \pi +\gamma$ for the three sets of parameters.
The ordering of those rates is the same as the ordering of $a_1$ radiative
decays: larger radiative decay gives larger results for $\pi+\rho
\rightarrow \pi +\gamma$.  Importantly, we point out that 
a $n$-times larger radiative decay does {\bf not} imply the 
$\pi+\rho \rightarrow \pi +\gamma$ rate to be larger by the same factor. 
This statement is accurate for the pure $a_1$ contribution to $\pi+\rho
\rightarrow \pi +\gamma$ (right panel of Fig. 1), but the total  $\pi+\rho
\rightarrow \pi +\gamma$ rate is given by a coherent sum of 
diagrams containing virtual $a_1$ and other meson species. 
Therefore, even if the photon yield from the $a_1$ diagrams with set
I is considerably larger than the corresponding contribution of set II 
(right panel Fig. 1), the result for set I becomes about four times 
larger than the result of set II, once we coherently add the other
contributions. Note that set II, which reproduces the hadronic 
phenomenology of the $a_1$ well, might be considered as a conservative 
estimate for the contributions of $\pi+\rho \rightarrow \pi +\gamma$, 
as it under-predicts the radiative decay.
To quantify this further, we adjusted parameters to reproduce the $a_1$
electromagnetic width (set III), yielding a photon rate for $\pi+\rho 
\rightarrow \pi +\gamma$ which is larger than that of set II by a factor 
of less than two. 
To have a rough estimate of the error associated with those
rates, one can simply consider the range spanned by the results 
of set II (good $D/S$, wrong
$\Gamma_{a_1 \rightarrow \pi +\gamma}$) and those of set III (wrong $D/S$,
good $\Gamma_{a_1 \rightarrow \pi +\gamma}$). The consequence of this exercise
is that set II is used here, and that we regard its inherent  uncertainty 
to be within a factor of two.  Note also that the MYM approach, being based 
on chiral U$(3)_L \times$ U$(3)_R$ symmetry, treats the interaction of 
non-strange mesons with strange
mesons in a coherent way. This is a desirable feature.  For the sake of
comparison, around $E_{\gamma}$=3 GeV, the result from set II is a factor 2
larger than that of Ref.~[\refcite{KLS91}], where the $a_1$ was not included.

\subsubsection{Form-factors and the strange sector}

\begin{figure}[ht!]
\psfig{figure=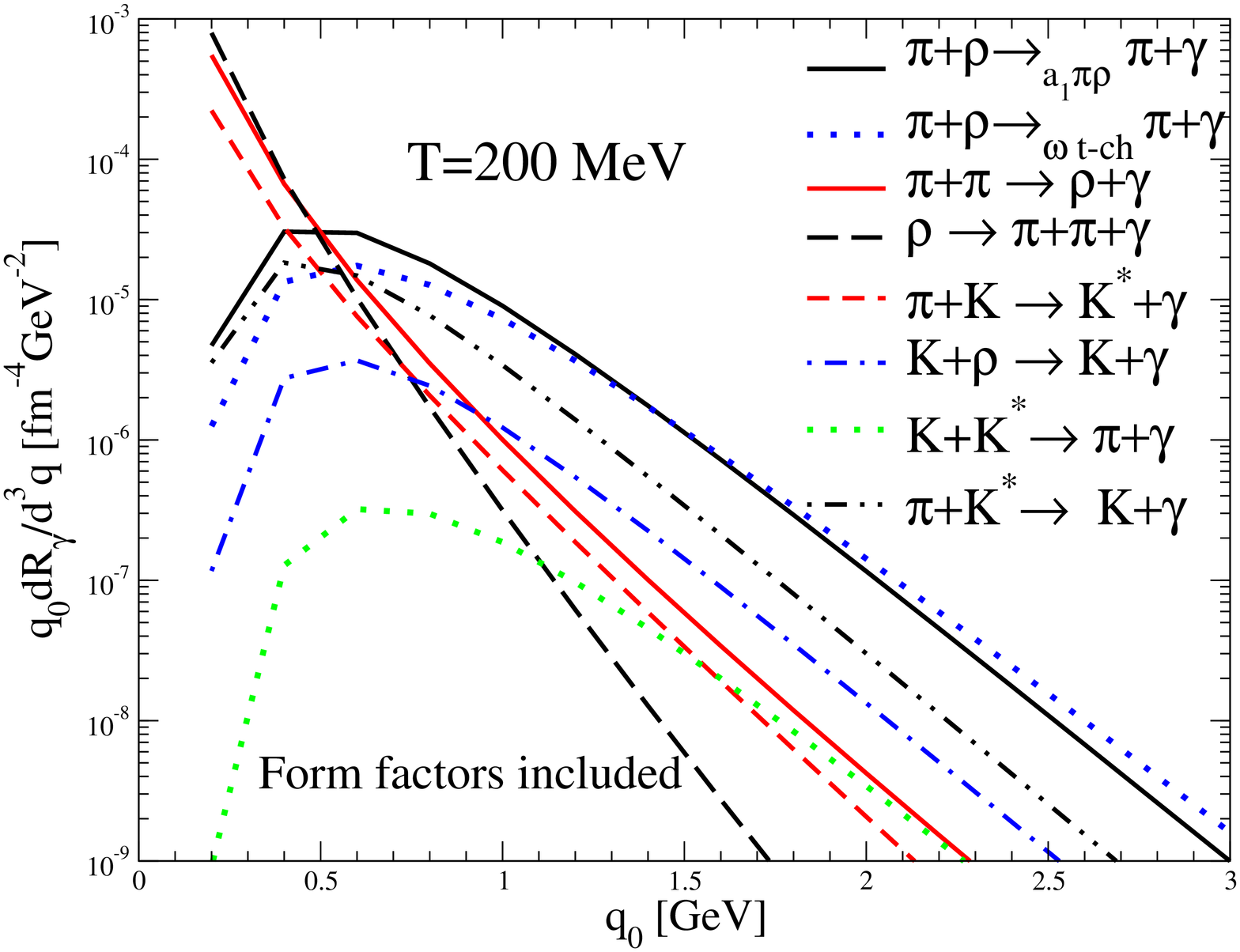,width=1.7in,angle=-90}
\psfig{figure=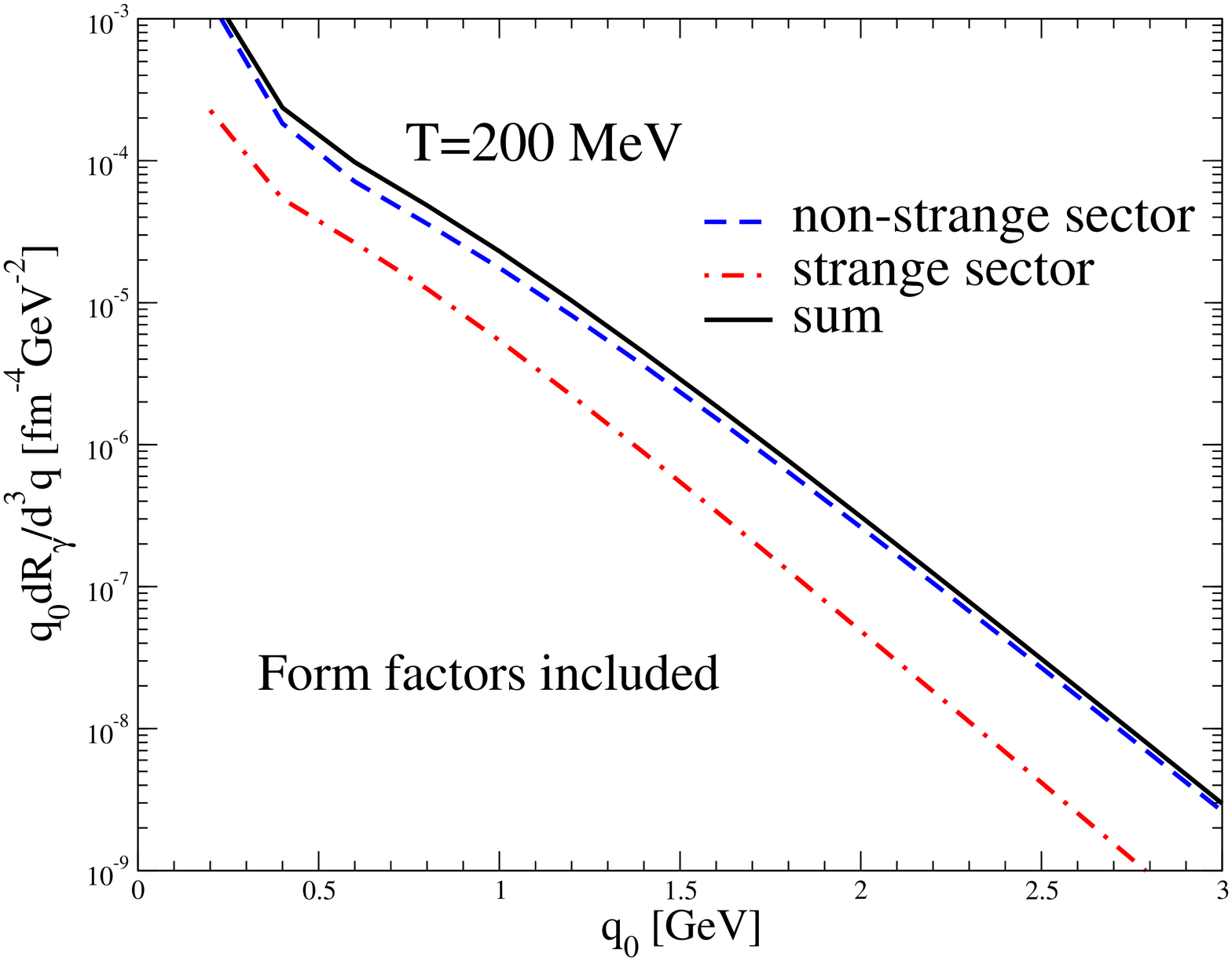,width=1.7in,angle=-90}
\caption{Strange and non-strange meson contributions to the production
rate of photons at T=200 MeV.}
\label{totrates}
\end{figure}

A conventional way to account for the finite size of mesons when using
effective models at high momentum transfer is to introduce
form factors.  To be consistent with
the procedure adopted before for dilepton production~\cite{RG99}, we
assume a standard dipole form for each hadronic vertex appearing in the
amplitudes,
\begin{equation}
F(t)=\left(\frac{2\Lambda^2}{2\Lambda^2-t}\right)^2 \ ,  
\label{ff_t}
\end{equation}
using a typical hadronic cutoff scale, $\Lambda=1$~GeV~\cite{RG99}.
We then approximate the four-momentum transfer in a given $t$-channel
exchange of meson $X$ by its average $\bar t$ according to~\cite{simon}
\begin{eqnarray}
\left(\frac{1}{m_{X}^2-\bar{t}}\right)^2
\left(\frac{2\Lambda^2}{2\Lambda^2-\bar{t}}\right)^8
=\frac{1}{4E^2}\int^{4E^2}_{0}
\frac{dt(2\Lambda^2)^8}{(m_{X}^2-t)^2(2\Lambda^2-t)^8} \ .
\label{ff_av}
\end{eqnarray}
This procedure allows to factorize
form factors from amplitudes which facilitates the task of
making the final expression gauge invariant.
Finally, using the MYM Lagrangian and the above form factor, we
calculate amplitudes for all possible (Born-) graphs for the reactions:
$X+Y\rightarrow Z+\gamma$, $\rho\rightarrow Y+Z+\gamma$ and
$K^*\rightarrow Y+Z+\gamma$. For $X$, $Y$, $Z$ every allowed
combination of $\rho, \pi,K^*,K$ mesons was considered.  
The results for the strange and non-strange
sectors are summarized in Fig.~2.  Overall, the total strange
contribution accounts for $\sim$ 25\% of the net contribution
around $q_0$=1 GeV, while this reduces to $\sim$ 15\% at $q_0$=3 GeV
(right panel of Fig. 2).  Parametrizations for all of those reactions are
given in Ref.~[\refcite{simon}]. It is of interest to point out that the rates
of Fig.~\ref{totrates} contain a contribution of an $\omega$ $t$-channel
exchange, in the case of $\pi \rho$ initial state. This process has been
found to contribute significantly, especially if the hadronic couplings are 
deduced from fits including hadronic form factors, as they should\cite{simon}.

\section{Other sources of photons in Heavy Ions Collisions}

We added baryonic sources to the mesonic contribution of the previous
section by evaluating in-medium vector spectral functions~\cite{RW99}
 at the photon point. An important consistency requirement is that
the spectral density used for photons needs to be the same as that employed
for the interpretation of dilepton measurements~\cite{ralfwa}. 
This requirement is fulfilled here, and
double-counting issues have received careful consideration.
In the case of dilepton production~\cite{ralfwa,ralf1}, baryons have
been found to be important sources of virtual photons. In the photon
sector, they manifest themselves in the low energy spectrum~\cite{simon}, 
$q_0 < $ 1 GeV. For the production of photons from the quark-gluon
plasma, we used the parametrized results of Ref.~[\refcite{guy1}], which
include the Laudau-Pomeranchuk-Migdal (LPM) effect~\cite{landau,migdal}
through a complete leading-order calculation~\cite{guy2,guy3}.

Finally, primordial nucleon-nucleon ($N$-$N$) collisions also contribute  
to the production of
direct photons; the pertinent spectra are given by~\cite{wong}
\begin{equation}
q_0 \frac{dN_\gamma^{prompt}}{d^3q} = q_0
\frac{d^3\sigma_\gamma^{pp}}{d^3q} \
 \frac{N_{coll}}{\sigma_{pp}^{in}}\ .
\end{equation}
$N_{coll}$ is the number of $N$-$N$ collisions, $\sigma_{pp}^{in}$ is
the total inelastic $N$-$N$ cross-section and $q_0
\frac{d^3\sigma_\gamma^{pp}}{d^3q}$ the differential cross-section to
produce photons, which includes intrinsic $k_T$ effects at the $N$-$N$
level~\cite{simon,dinesh}.  An additional broadening, attributable to the
Cronin effect~\cite{cronin}, is obtained by folding over a Gaussian
transverse momentum distribution~\cite{simon,dumitru}:
\begin{equation}
f (k_T) = \frac{1}{\pi \langle \Delta k_T^2 \rangle} \
{\rm e}^{-k_T^2/\langle \Delta k_T^2 \rangle} \ .
\end{equation}
The amount of broadening has been estimated by fitting $p$-$A$ data, 
and extrapolating to $A$-$A$ collisions~\cite{simon}. One then obtains 
$\langle\Delta k_T^2\rangle\sim$ 0.2-0.3 GeV$^2$.

\section{Comparison with experiment and predictions}

Our total photon production from QGP and hadronic phase is a
convolution of the production rate of the previous sections with a
fireball evolution model~\cite{RW99,RS00} adjusted to reproduce 
observed flow velocities and hadron ratios. On top of this, we add the 
photon yield from primordial $N$-$N$ collisions.  With an initial
temperature of $T_i=$ 205 MeV, our results (left panel Fig. 3) are
consistent with WA98 data~\cite{wa98} over the entire spectrum.  The
high and low $q_t$ regions are dominated by primordial $N$-$N$
collisions and hadronic phase, respectively.  
The QGP appears to be sub-dominant over all of the energy range.
However, our results for RHIC (right panel of Fig. 3) show that 
there is a window, 1 GeV $<q_t<3$ GeV, through which the QGP shines. 
The Cronin effect has been neglected for RHIC, as it is expected to be less
important than at the SPS~\cite{dumitru,adler}.

\begin{figure}[ht!]
\begin{center}
\psfig{file=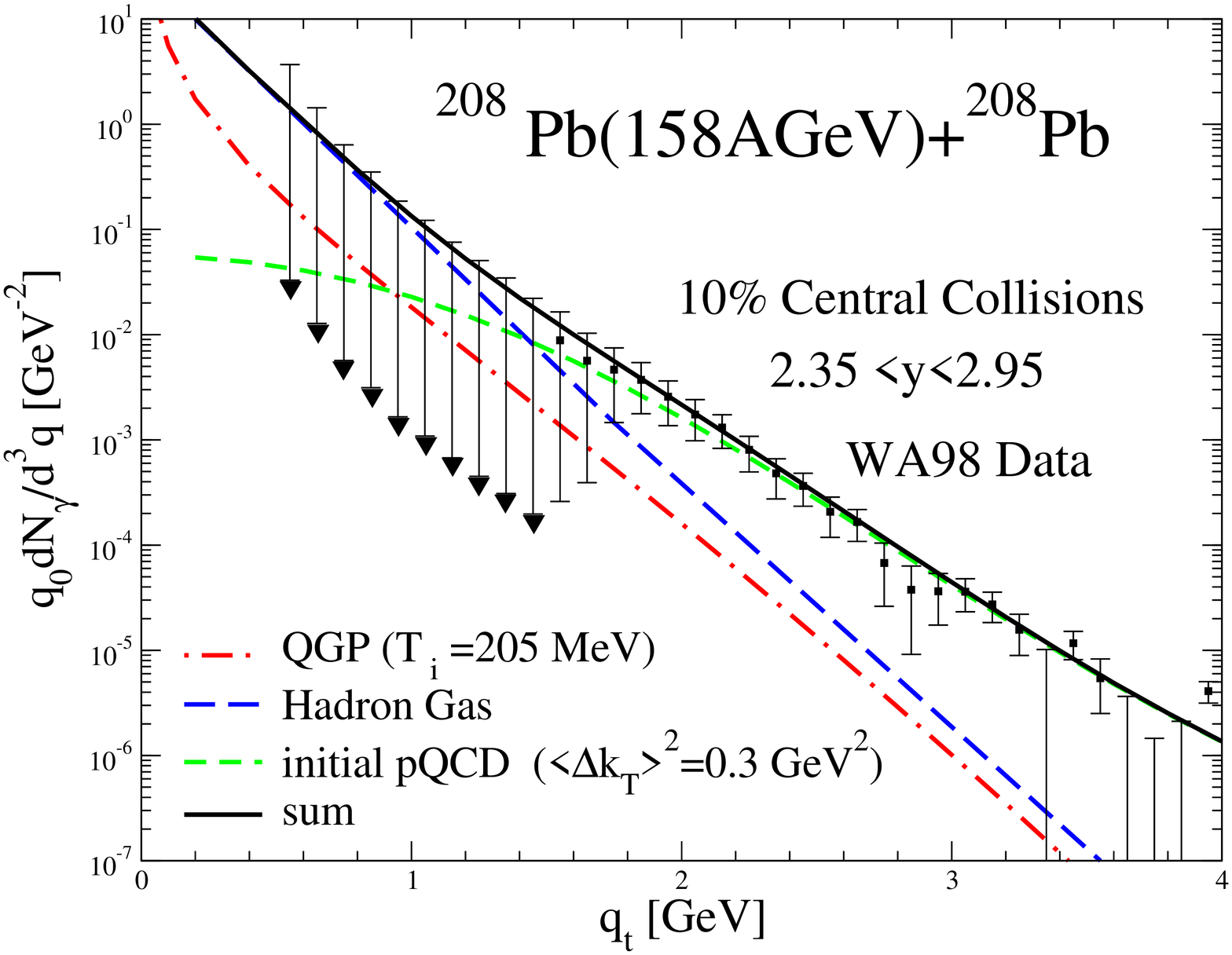,width=4.3cm,angle=-90}
\psfig{file=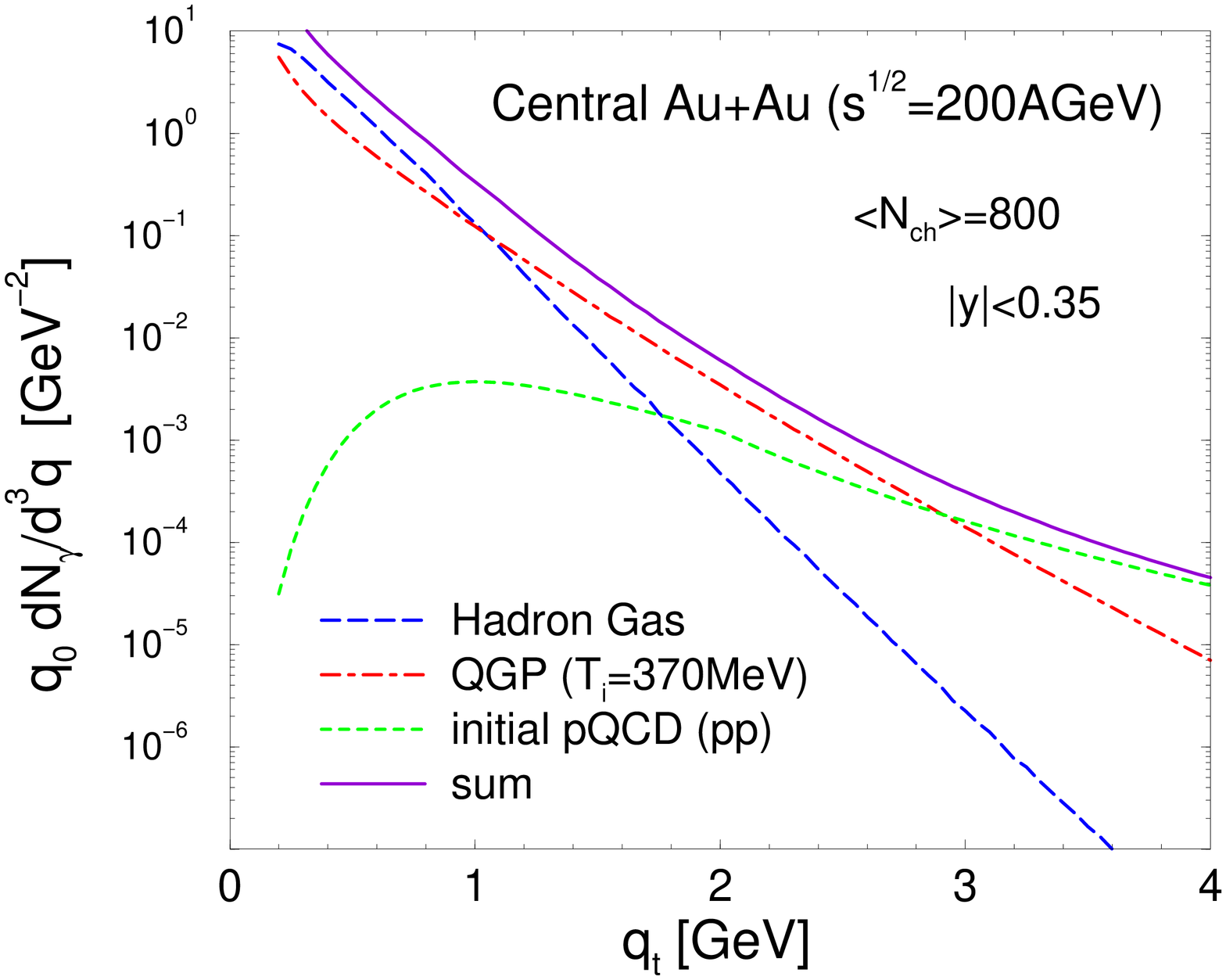,width=4.4cm,angle=-90}
\caption{Left panel: thermal plus prompt photon spectra compared to
data from WA98~\protect\cite{wa98}
for central $Pb$+$Pb$ collisions at SPS.  Right panel: photon spectra
at RHIC. }
\end{center}
\end{figure}

\section{CONCLUSIONS}

We have shown results for the production rate of real photons calculated 
within a chiral Lagrangian based on a Massive Yang-Mills approach.  
The pertinent hadronic rate was supplemented
by the leading order QCD calculation for QGP to evaluate spectra in 
central $Pb$-$Pb$ collisions at SPS using a thermal fireball evolution 
model.  WA98 data are reasonably well reproduced once primordial 
$N$-$N$ collisions are also included with a Cronin effect estimated 
from $p$-$A$ data. Strange particles are found to have a non-negligible
contribution to the photon production rate, $\sim$ 15-25 \% of the
total hadronic contribution.
The QGP contribution is found to be around $\sim$ 20-30\% of the thermal 
yield, implying that the WA98 data can be interpreted, to a large
extent, within a hadronic framework. 
At RHIC, the QGP contribution becomes dominant for intermediate photon 
energies of around 2~GeV, which provides a promising prospect for the 
measurement of electromagnetic radiation from relativistic nuclear collisions.  


\section*{Acknowledgments}

This work was supported in part by
the Natural Sciences and Engineering Research Council of Canada, and in
part
by the Fonds Nature et Technologies of Quebec.

\end{document}